\documentclass[12pt]{article}
\usepackage{amsmath,amssymb,amsfonts,graphicx}

\begin{document}
\thispagestyle{empty}
\renewcommand{\refname}{References}

\title{\bf Optical theorem for Aharonov-Bohm scattering}
\author{Yu.A.~Sitenko$^{1}$ and N.D.~Vlasii$^{1,2}$}
\date{}

\maketitle

\begin{center}
$^{1}$Bogolyubov Institute for Theoretical Physics, \\
National Academy of Sciences of Ukraine, \\ 14-b Metrologichna
Str., Kyiv, 03680, Ukraine \\
$^{2}$Physics Department, Taras Shevchenko National University of Kyiv, \\
64 Volodymyrska str., Kyiv, 01601, Ukraine
\end{center}

\begin{abstract}
Quantum-mechanical scattering off a magnetic vortex is considered,
and the optical theorem is derived. The vortex core is assumed to be
impermeable to scattered particles, and its transverse size is taken
into account. We show that the scattering Aharonov-Bohm effect is
independent of the choice of boundary conditions from the variety of
the Robin ones. The behaviour of the scattering amplitude in the
forward direction is analyzed, and the persistence of the Fraunhofer
diffraction in the short-wavelength limit is shown to be crucial for
maintaining the optical theorem in the quasiclassical limit.
\end{abstract}


\bigskip

\begin{center}
\end{center}

\bigskip
\medskip

\section{Introduction}

Probability conservation and the unitarity of the scattering matrix
are the basic elements of quantum theory, which have significant
physical consequences. One of them is the optical theorem relating
the imaginary part of the scattering amplitude in the forward
direction to the total cross section of the interaction processes.
Quantum-mechanical scattering of a charged particle off an
impermeable magnetic vortex is studied for a more than half a
century, starting from the seminal paper of Aharonov and Bohm
\cite{Aha}. The theory of this process has been successfully
confirmed in experiments, promising important practical
applications, see review \cite{Ton}. However, some theoretical
issues still remain unclear, and among them the question on how to
formulate the optical theorem for the case of the Aharonov-Bohm
scattering. Several authors \cite{Sak,Ara} addressed this problem
but without any decisive conclusion (see also \cite{Som}); they
traced the encountered difficulties either to a subtlety in the
choice of an incident wave \cite{Sak}, or to a divergent behaviour
of the scattering amplitude in the forward direction, which needs
yet unspecified regularization \cite{Ara}. In our opinion, the
following two circumstances should be taken into consideration: a)
the long-range nature of interaction of a scattering particle with
the magnetic vortex, and b) the nonvanishing transverse size of the
magnetic vortex. Due to the first circumstance, although the
unitarity of the $S$-matrix is undoubted, the standard scattering
theory which is applicable to the case of short-range interactions
has to be modified, resulting in a rather unexpected form of the
optical theorem. The second circumstance signifies that the limit of
the vanishing transverse size of the vortex is an undue idealization
which has to be avoided as physically irrelevant. Then, provided
that points a) and b) are taken into account, the above problem can
be treated properly and resolved, as it is shown in the present
paper.

The magnetic field configuration in the form of an infinitely long
vortex possesses cylindrical symmetry. The $S$-matrix in the
cylindrically symmetric case takes form
\begin{equation}
S(k,\,\varphi,k_z;\,k',\,\varphi',k'_z)=\biggl[I(k,\,\varphi;\,k',\,\varphi')
+\delta(k-k')\frac{\rm i}{\sqrt{2\pi k}}
f(k,\,\varphi-\varphi')\biggr] \delta(k_z - k'_z),\label{eq1}
\end{equation}
where the symmetry axis coincides with the $z$-axis, ${\rm
I}(k,\,\varphi;\,k',\,\varphi')$ is the unity matrix in polar
coordinates in two-dimensional space, and $f(k,\,\varphi-\varphi')$
is the scattering amplitude. The condition of the unitarity of the
$S$-matrix, $S^\dag S=SS^\dag={\rm I}$, results, in a conventional
way, in the optical theorem
\begin{equation}
2\sqrt{\frac{2\pi}{k}}{\rm Im}f(k,\,0)=\sigma, \label{eq2}
\end{equation}
where $\sigma$ is the total cross section per unit length along the
symmetry axis, and, thus, $\sigma$ has the dimension of length. We
shall prove that, namely in the case of the Aharonov-Bohm
scattering, the optical theorem takes a form which is different from
the conventional one given by (2). Our consideration is based on
earlier works \cite{Be,Rui,Sou,Jac,Si2,Si5} where important results
concerning the scattering wave function, $S$-matrix and scattering
amplitude have been obtained.

In the next section we review an auxiliary problem of scattering off
an impermeable tube; the role of the Fraunhofer diffraction and the
appropriate forward peak is exposed. In Section 3 we consider the
same problem for the case when the tube is filled with the magnetic
flux lines, i.e. for the case of an impermeable magnetic vortex; the
main results are derived here. Summary and discussion of the results
are given in Section 4. We relegate the details of calculation of
the scattering amplitude and the total cross section in the
quasiclassical limit to Appendices A and B.

\section{Scattering off an impermeable tube}
A plane wave propagating in the direction which is orthogonal to the
$z$-axis can be presented as
\begin{eqnarray}
\psi_{\bf k}^{(0)}({\bf r})&=&e^{{\rm i}kr\cos\varphi}=\sum\limits_{n\in\mathbb{Z}}e^{{\rm i}n\varphi}
e^{\frac{{\rm i}}{2}|n|\pi}J_{|n|}(kr)=\nonumber \\
&=&\frac 12\sum\limits_{n\in\mathbb{Z}}e^{in\varphi}e^{\frac{\rm i}2|n|\pi}
\left[H_{|n|}^{(1)}(kr)+H_{|n|}^{(2)}(kr)\right],\label{eq3}
\end{eqnarray}
where ${\bf r}$ and ${\bf k}$ are the two-dimensional vectors,
$\varphi$ is the angle between them, $J_\alpha(u)$,
$H_\alpha^{(1)}(u)$ and $H_\alpha^{(2)}(u)$ are the Bessel, first-
and second-kind Hankel functions of order $\alpha$, $\mathbb{Z}$ is
the set of integer numbers. At $r\rightarrow\infty$, using the
appropriate asymptotics of the Hankel functions, one gets
\begin{eqnarray}
\psi_{\bf k}^{(0)}({\bf
r})\,\,=\!\!\!\!\!\!\!\!_{_{r\rightarrow\infty}}\,\frac{1}{\sqrt{2\pi
kr}} \sum\limits_{n\in\mathbb{Z}}e^{{\rm i}n\varphi}\,e^{\frac {\rm
i}2|n|\pi}\left[e^{{\rm i}\left(kr-\frac12|n|\pi- \frac14\pi\right)}+\right.\nonumber \\
\left.+e^{-{\rm i}\left(kr-\frac 12|n|\pi-\frac 14\pi\right)}
\right]=\sqrt{\frac{2\pi}{kr}}e^{{\rm i}(kr-\pi/4)}\Delta(\varphi)+
\sqrt{\frac{2\pi}{kr}}e^{-{\rm
i}(kr-\pi/4)}\Delta(\varphi-\pi),\label{eq4}
\end{eqnarray}
where
\begin{equation}
\Delta(\varphi)=\frac 1{2\pi}\sum\limits_{n\in \mathbb{Z}}e^{{\rm
i}n\varphi}\label{eq5}
\end{equation}
is the delta-function for the azimuthal angle,
$\Delta(\varphi+2\pi)=\Delta(\varphi)$. Thus, we see that the plane
wave passing through the origin ($r=0$) can be naturally interpreted
at large distances from the origin as a superposition of two
cylindrical waves: the diverging one, $e^{{\rm i}kr}$, in the
forward, $\varphi=0$, direction and the converging one, $e^{-{\rm
i}kr}$, from the backward, $\varphi=\pi$, direction.

Now, let us place an obstacle in the form of an opaque tube along
the $z$-axis. If the wave function obeys the Dirichlet boundary
condition at the edge of the tube,
\begin{equation}
\left.\psi_{\bf k}(\bf r)\right|_{r=r_c}=0,\label{eq6}
\end{equation}
then, instead of (3), we get
\begin{equation}
\psi_{\bf k}({\bf r})=\sum\limits_{n\in\mathbb{Z}}e^{{\rm i}n\varphi}e^{\frac {\rm i}2|n|\pi}
\left[J_{|n|}(kr)-\frac{J_{|n|}(kr_c)}{H_{|n|}^{(1)}(kr_c)}H_{|n|}^{(1)}(kr)\right].\label{eq7}
\end{equation}
At large distances from the origin, $r\gg k^{-1}$, we get
\begin{equation}
\psi_{\bf k}({\bf r})\,\,=\!\!\!\!\!\!\!\!_{_{kr\gg 1}}\,-\frac{1}{\sqrt{2\pi kr}}
e^{{\rm i}(kr-\pi/4)}\sum\limits_{n\in\mathbb{Z}}e^{{\rm i}n\varphi}\frac{H_{|n|}^{(2)}(kr_c)}
{H_{|n|}^{(1)}(kr_c)}
+\sqrt{\frac{2\pi}{kr}}e^{-{\rm i}(kr-\pi/4)}\Delta(\varphi-\pi).\label{eq8}
\end{equation}
The sum over $n$ in (8) yields the forward angular delta-function,
$\Delta(\varphi)$, in the case of long wavelengths, $k\rightarrow
0$. In the opposite, short-wavelength, limit, $k\rightarrow\infty$,
when $1\ll kr_c<kr$, one can get by substituting the large-argument
asymptotics of the Hankel functions in (8):
\begin{equation}
\psi_{\bf k}({\bf r})\,\,\,\,\,=\!\!\!\!\!\!\!\!\!\!\!\!\!_{_{kr>kr_c\gg 1}}\,-\sqrt{\frac{2\pi}{kr}}
e^{{\rm i}\pi/4}\left[e^{{\rm i}k(r-2r_c)}-e^{-{\rm i}kr}\right]\Delta(\varphi-\pi).\label{eq9}
\end{equation}
This result which is actually given in the monograph of Morse and
Feshbach \cite{Mor} is quite understandable from the classical point
of view: the obstacle forms a shadow in the forward direction, which
is not accessible to waves, and, thus, both diverging and converging
cylindrical waves are in the backward direction. However, this
conclusion is wrong.

To find a loophole in the arguments leading to (9), one has to note
a property of the asymptotical behaviour of the Bessel function at
large values of its argument: it vanishes effectively when its order
exceeds its large argument. Really, using integral representation
(see, e.g. \cite{Abra})
$$
J_{|n|}(u)=\frac{1}{2\pi}\int\limits_{-\pi}^{\pi}d\theta\,\exp\left[{\rm
i}(|n|\theta-u\sin\theta)\right],
$$
one notes that the integrand at large $|n|$ is vigorously
oscillating and its mean value is small almost everywhere with the
exception of points where the phase is stationary. This means that
the prevailing contribution to the integral in the case of $|n|\gg
1$ is given by the vicinity of the point where $\cos\theta=|n|/u$,
and, consequently, the integral is vanishingly small in the case of
$|n|/u\gg 1$. The more is $u$, the more abrupt is the decrease of
the integral when $|n|$ exceeds $u$ (see, e.g., \cite{New}).

Therefore, in (7) at $kr>kr_c\gg 1$, the sum containing
$J_{|n|}(kr)$ is cut at $|n|=kr$, while the sum containing
$J_{|n|}(kr_c)$ is cut at $|n|=kr_c$. Instead of (9), we get the
correct expression:
\begin{eqnarray}
\psi_{\bf k}({\bf r})\,\,\,\,\,=\!\!\!\!\!\!\!\!\!\!\!\!\!_{_{kr>kr_c\gg 1}}\,\sqrt{\frac{2\pi}{kr}}
e^{{\rm i}(kr-\pi/4)}\left[\Delta_{kr}(\varphi)-\Delta_{kr_c}(\varphi)\right]+
\nonumber \\ +\sqrt{\frac{2\pi}{kr}}e^{{\rm i}\pi/4}\left[e^{-{\rm i}kr}\Delta_{kr}(\varphi-\pi)
-e^{{\rm i}k(r-2r_c)}\Delta_{kr_c}(\varphi-\pi)\right],\label{eq10}
\end{eqnarray}
where
\begin{equation}
\Delta_x(\varphi)=\frac{1}{2\pi}\sum\limits_{|n|\leq x}e^{{\rm i}n\varphi}\label{eq11}
\end{equation}
is the regularized (smoothed) angular delta-function,
\begin{equation}
\lim\limits_{x\rightarrow\infty}\Delta_x(\varphi)=\Delta(\varphi),\qquad \Delta_x(0)=\frac{x}{\pi}.\label{eq12}
\end{equation}
We see that a cancellation of the diverging wave in the forward
direction is not complete, as well as is that between the diverging
and the converging waves in the backward direction; the complete
cancellation is achieved at $r=r_c$ only, that is consistent with
condition (6). In general, the diverging wave in the forward
direction is suppressed by factor $1-r_cr^{-1}$, as compared to the
case when the obstacle is absent. Thus, contrary to the classical
anticipations, the wave penetrates to the region behind the obstacle
even in the case when the wavelength is much less than the
transverse size of the obstacle, $kr_c\gg 1$.

Turning now from the qualitative analysis to the quantitative one,
we note, first, from (7) that the asymptotics of the wave function
at large distances from the obstacle is
\begin{equation}
\psi_{\bf k}({\bf r})=\psi_{\bf k}^{(0)}({\bf r})+f(k,\,\varphi)\frac{e^{{\rm i}(kr+\pi/4)}}{\sqrt{r}}+O(r^{-3/2}),
\label{eq13}
\end{equation}
where
\begin{equation}
f(k,\,\varphi)={\rm i}\sqrt{\frac{2}{k\pi}}\sum\limits_{n\in\mathbb{Z}}
e^{{\rm i}n\varphi}\frac{J_{|n|}(kr_c)}{H_{|n|}^{(1)}(kr_c)}\label{eq14}
\end{equation}
is the scattering amplitude which enters $S$-matrix (1), while the
unity matrix there is evidently
\begin{equation}
I(k,\,\varphi;\,k',\,\varphi')=\frac 1k\delta(k-k')\Delta(\varphi-\varphi').\label{eq15}
\end{equation}
It should be noted that wave function (7), apart from condition (6),
satisfies also condition
\begin{equation}
\lim\limits_{r\rightarrow \infty}e^{{\rm i}kr}\psi_{\bf k}({\bf r})|_{\varphi=\pm\pi}=1,
\label{eq16}
\end{equation}
signifying that the incident wave comes from the far left; the
forward direction is $\varphi=0$, and the backward direction is
$\varphi=\pm \pi$.

The $S$-matrix is unitary:
\begin{eqnarray}
\int\limits_{-\infty}^{\infty}d\,k_z\int\limits_{0}^{\infty}dk\,k\int\limits_{-\pi}^{\pi}d\varphi\,
S^*(k,\,\varphi,\,k_z;\,k',\,\varphi',\,k_z')S(k,\,\varphi,\,k_z;\,k'',\,\varphi'',\,k_z'')=\nonumber \\=\frac{1}{k'}\delta
(k'-k'')\Delta(\varphi'-\varphi'')\delta(k_z'-k_z''),\label{eq17}
\end{eqnarray}
and the latter relation can be recast into the form
\begin{equation}
\frac 1{\rm i}\sqrt{\frac{k}{2\pi}}\left[f(k,\,\varphi'-\varphi'')-f^*(k,\,\varphi''-\varphi')\right]
=\frac{k}{2\pi}\int\limits_{-\pi}^{\pi}d\varphi\,f^*(k,\,\varphi-\varphi')f(k,\,\varphi-\varphi''),\label{eq18}
\end{equation}
in particular, at $\varphi'=\varphi''=0$ one gets optical theorem
(2) with $\sigma=\int\limits_{-\pi}^{\pi}d\varphi|f(k,\,\varphi)|^2$
being the total cross section for elastic scattering. Although (18)
is valid for all wavelengths, it relates vanishingly small
quantities in the long-wavelength limit, whereas in the
short-wavelength limit it relates extremely large quantities.

Really, the scattering amplitude in the long-wavelength limit is
estimated as
\begin{eqnarray}
f(k,\,\varphi)\!=\!-\sqrt{\frac{\pi}{2k}}|\ln (kr_c)|^{-1}\left[1\!+\!\left(\gamma\!-\!{\rm i}\frac \pi 2\right)
|\ln(kr_c)|^{-1}\right]\!+\!k^{-1/2}O\left[|\ln(kr_c)|^{-3}\right]\!,\nonumber \\
 kr_c\ll 1\label{eq19}
\end{eqnarray}
($\gamma$ is the Euler constant), and (18) is the relation between
quantities of order $O\left[|\ln(kr_c)|^{-2}\right]$. The scattering
amplitude in the short-wavelength limit is
\begin{eqnarray}
f(k,\,\varphi)={\rm i}\sqrt{\frac{2\pi}{k}}\Delta_{kr_c}(\varphi)-
\sqrt{\frac{r_c}{2}|\sin(\varphi/2)|}\exp\left\{-2{\rm i}kr_c|\sin(\varphi/2)|\right\}e^{-{\rm i}\pi/4}, \nonumber \\
kr_c\gg 1.\label{eq20}
\end{eqnarray}
The first term in (20) represents the forward peak of the Fraunhofer
diffraction on the obstacle, whereas the second term describes the
reflection from the obstacle according to the laws of geometric
(ray) optics. Hence, (18) in this case is the relation between
quantities of order $O(kr_c)$ at $\varphi'=\varphi''$ or of order
$O(\sqrt{kr_c})$ at $\varphi'\neq \varphi''$, and optical theorem
(2) is the relation between finite quantities of order $r_c$. It
should be noted that the Fraunhofer diffraction is crucial for
ensuring the optical theorem in the short-wavelength limit, since
the second term in (20) vanishes at $\varphi=0$.

\section{Scattering off an impermeable magnetic vortex}

Let us consider scattering of a charged particle off an obstacle in
the form of an impermeable tube which is filled with magnetic field
of total flux $\Phi$. The particle wave function obeys conditions
(6) and (16); an additional circumstance is that the Schr\"{o}dinger
hamiltonian out of the tube is no longer free but takes form
\begin{equation}
H=-\frac{\hbar^2}{2m}\left[\partial_r^2+\frac 1r\partial_r+\frac{1}{r^2}\left(\partial_\varphi-
{\rm i}\frac{\Phi}{\Phi_0}\right)^2+\partial_z^2\right],\label{eq21}
\end{equation}
where $\Phi_0=2\pi\hbar ce^{-1}$ is the London flux quantum.
Therefore, the scattering wave solution in this case is (cf. (7))
\begin{equation}
\psi_{\bf k}({\bf r})=\sum\limits_{n\in \mathbb{Z}}e^{{\rm i}n\varphi}e^{{\rm i}(|n|-\frac 12|n-\mu|)\pi}
\left[J_{|n-\mu|}(kr)-\frac{J_{|n-\mu|}(kr_c)}{H_{|n-\mu|}^{(1)}(kr_c)}H_{|n-\mu|}^{(1)}(kr)\right],\label{eq22}
\end{equation}
where $\mu=\Phi\Phi_0^{-1}$.

In the long-wavelength limit, we get the asymptotics
\begin{eqnarray}
\psi_{\bf k}({\bf r})\,\,\,\,=\!\!\!\!\!\!\!\!\!\!\!_
{_{{kr\gg 1}\atop{kr_c\ll 1}}}
\sqrt{\frac{2\pi}{kr}}\,e^{{\rm i}(kr-\pi/4)}\left[\cos(\mu\pi)\Delta(\varphi)-\sin(\mu\pi)
\Gamma^{(\nu)}(\varphi)\right]+ \nonumber \\
+\sqrt{\frac{2\pi}{kr}}\,e^{-{\rm i}(kr-\pi/4)}\Delta(\varphi-\pi),\label{eq23}
\end{eqnarray}
where $\nu=[\![\mu]\!]$ and $[\![u]\!]$ denotes the integer part of
$u$ (i.e. the integer which is less than or equal to $u$),
\begin{equation}
\Gamma^{(\nu)}(\varphi)=\frac{1}{2\pi{\rm i}}\sum\limits_{n\in\mathbb{Z}}{\rm sgn}(n-\mu)e^{{\rm i}n\varphi}\label{eq24}
\end{equation}
and ${\rm sgn}(u)=\left\{\begin{array}{cc}
                             1, & u>0 \\
                             -1,  & u<0
                           \end{array}
\right.$. In the short-wavelength limit, we get the asymptotics
\begin{eqnarray}
\psi_{\bf k}({\bf r})\,\,\,\,\,=\!\!\!\!\!\!\!\!\!\!\!\!_
{_{kr>kr_c\gg 1}}
\sqrt{\frac{2\pi}{kr}}\,e^{{\rm i}(kr-\pi/4)}\left\{\cos(\mu\pi)\left[\Delta_{kr}^{(\nu)}(\varphi)-\Delta_{kr_c}^{(\nu)}
(\varphi)\right]-\right.\nonumber \\
\left.-\!\sin(\mu\pi)\!\left[\Gamma^{(\nu)}_{kr}(\varphi)\!-\!\Gamma_{kr_c}^{(\nu)}(\varphi)\right]\!\right\}
\!+\!\sqrt{\frac{2\pi}{kr}}\,e^{{\rm i}\pi/4}\left[e^{-{\rm i}kr}\Delta_{kr}^{(\nu)}
(\varphi\!-\!\pi)\!-\!e^{{\rm i}k(r-2r_c)}
\Delta_{kr_c}^{(\nu)}(\varphi\!-\!\pi)\right],\label{eq25}
\end{eqnarray}
where
\begin{equation}
\Delta_x^{(\nu)}(\varphi)=\frac{1}{2\pi}\sum\limits_{|n-\mu|\leq x}e^{in\varphi}, \qquad
\Gamma_x^{(\nu)}(\varphi)=\frac{1}{2\pi{\rm i}}\sum\limits_{|n-\mu|\leq x}{\rm sgn}(n-\mu)e^{{\rm i}n\varphi};\label{eq26}
\end{equation}
note that $\Delta_x^{(\nu)}(\varphi)$ can be regarded as a
regularization for $\Delta(\varphi)$ (5), since
$\Delta_x^{(\nu)}(\varphi)$, as well as $\Delta_x(\varphi)$ (11),
satisfies conditions (12). It should be also noted that, since it is
a merely qualitative analysis, the long-wavelength asymptotics of
the wave function can be estimated as well as (23) with
$\Delta_{kr}^{(\nu)}$ and $\Gamma_{kr}^{(\nu)}$ substituted for
$\Delta$ and $\Gamma^{(\nu)}$.

Thus, we conclude that, both in the long- and short-wavelength
limits the particle wave penetrating in the forward direction behind
the obstacle depends periodically as cosine on the enclosed magnetic
flux with the period equal to the London flux quantum. This periodic
dependence, as well as the sine periodic dependence of the reflected
wave in other directions, is due to the fact that the interaction
with the scatterer is neither of the potential type, nor of the
sufficient decrease at large distances from the scatterer, see
hamiltonian (21).

Turning now from the qualitative analysis to the quantitative one,
we note, first, that, owing to the long-range nature of interaction,
the unity matrix in (1) is distorted:
\begin{equation}
I(k,\,\varphi;\,k',\,\varphi')=\cos(\mu\pi)\frac 1k\delta(k-k')\Delta(\varphi-\varphi'),\label{eq27}
\end{equation}
while the scattering amplitude is
\begin{equation}
f(k,\,\varphi)=f_0(k,\,\varphi)+f_c(k,\,\varphi),\label{eq28}
\end{equation}
where
\begin{equation}
f_0(k,\,\varphi)={\rm i}\sqrt{\frac{2\pi}{k}}\sin(\mu\pi)\Gamma^{(\nu)}(\varphi)\label{eq29}
\end{equation}
and
\begin{equation}
f_c(k,\,\varphi)={\rm i}\sqrt{\frac{2}{k\pi}}\sum\limits_{n\in\mathbb{Z}}
e^{{\rm i}n\varphi}e^{{\rm i}(|n|-|n-\mu|)\pi}\frac{J_{|n-\mu|}(kr_c)}{H^{(1)}_{|n-\mu|}(kr_c)}.\label{eq30}
\end{equation}
The asymptotics of the wave function at large distances from the
vortex is
\begin{equation}
\psi_{\bf k}({\bf r})=\psi^{(0)}_{\bf k}({\bf r})e^{{\rm i}\mu[\varphi-{\rm sgn}(\varphi)\pi]}+
f(k,\,\varphi)\frac{e^{{\rm i}(kr+\pi/4)}}{\sqrt{r}}+O(r^{-3/2}),\label{eq31}
\end{equation}
where it is implied that $-\pi<\varphi<\pi$.

From now on, we extend our consideration to involve a more general
boundary condition for the wave function at the edge of the
impermeable magnetic vortex. Namely, we impose the Robin boundary
condition,
\begin{equation}
\left.\left\{\left[\cos(\rho\pi)+r_c\sin(\rho\pi)\frac{\partial}{\partial r}\right]\psi_{\bf k}({\bf r})\right\}
\right|_{r=r_c}=0,\label{eq32}
\end{equation}
then (22) is changed to
\begin{equation}
\psi_{\bf k}({\bf r})=\sum_{n\in \mathbb{Z}}e^{{\rm i}n\varphi}e^{{\rm i}\left(|n|-\frac 12|n-\mu|\right)\pi}
\biggl[J_{|n-\mu|}(kr)-\Upsilon_{|n-\mu|}^{(\rho)}(kr_c)H_{|n-\mu|}^{(1)}(kr)\biggr],
\label{eq33}
\end{equation}
and (30) is changed to
\begin{equation}
f_c(k,\,\varphi)={\rm i}\sqrt{\frac{2}{k\pi}}\sum\limits_{n\in\mathbb{Z}}e^{{\rm i}n\varphi}
e^{{\rm i}(|n|-|n-\mu|\pi)}\Upsilon_{|n-\mu|}^{(\rho)}(kr_c),\label{eq34}
\end{equation}
where
\begin{equation}
\Upsilon_{\alpha}^{(\rho)}(u)=\frac{\cos(\rho\pi)J_{\alpha}(u)+\sin(\rho\pi)u\frac{d}{du}J_\alpha(u)}
{\cos(\rho\pi)H_{\alpha}^{(1)}(u)+\sin(\rho\pi)u\frac{d}{du}H_{\alpha}^{(1)}(u)}.\label{eq35}
\end{equation}
Hence, $\rho=0$ corresponds to the Dirichlet condition (perfect
conductivity of the boundary), see (6), and $\rho=1/2$ corresponds
to the Neumann condition (absolute rigidity of the boundary),
\begin{equation}
\left.\left[\frac{\partial}{\partial r}\psi_{\bf k}({\bf r})\right]\right|_{r=r_c}=0.\label{eq36}
\end{equation}

\subsection{Optical theorem}

The unitarity condition for the $S$-matrix, (17), is certainly valid
for all wavelengths. However, due to the long-range nature of
interaction, the consequent condition in terms of the scattering
amplitude takes forms which differ from (18).

The long-wavelength limit, $k\rightarrow 0$, is the same as the
$r_c\rightarrow 0$ limit corresponding to the idealized case of a
singular vortex of zero thickness. Amplitude $f_c$ (34) can be
neglected in this case, and the $S$-matrix unitarity condition
involves amplitude $f_0$ (29) only. In view of (27) and relation
\begin{equation}
\Gamma^{(\nu)}(\varphi)+\left[\Gamma^{(\nu)}(-\varphi)\right]^*=0, \label{eq37}
\end{equation}
we get, instead of (18), the following relation:
\begin{equation}
\sin^2(\mu\pi)\Delta(\varphi'-\varphi'')=\frac{k}{2\pi}\int\limits_{-\pi}^{\pi}d\varphi\,
f_0^*(k,\,\varphi-\varphi')f_0(k,\,\varphi-\varphi'').\label{eq38}
\end{equation}
Thus, we see that the optical theorem which should be derived from
(38) by putting $\varphi'=\varphi''=0$ is hardly informative, being
a relation between infinite quantities, $\Delta(0)$.

The failure with the optical theorem for the Aharonov-Bohm
scattering in the long-wavelength limit is due to an unphysical
idealization inherent in the treatment of this case. As long as the
transverse size of the vortex is taken into account, the optical
theorem emerges as a relation between finite quantities. Really,
retaining $f_c$ in the unitarity relation (17), we get
\begin{eqnarray}
\frac 1{\rm i}\sqrt{\frac{k}{2\pi}}\biggl\{\cos(\mu\pi)\left[f_c(k,\,\varphi'-\varphi'')-
f^*_c(k,\,\varphi''-\varphi')\right]+\biggr. \nonumber \\ \biggl.+\sin(\mu\pi)\int\limits_{-\pi}^{\pi}d\varphi
[\Gamma^{(\nu)}(\varphi'-\varphi)f_c(k,\,\varphi-\varphi'')+\biggr. \nonumber \\  \biggl.+
f^*_c(k,\,\varphi-\varphi')\Gamma^{(\nu)}(\varphi-\varphi'')]\biggr\}
=\frac{k}{2\pi}\int\limits_{-\pi}^{\pi}d\varphi\,f_c^*(k,\,\varphi-\varphi')
f_c(k,\,\varphi-\varphi''),\label{eq39}
\end{eqnarray}
which at $\varphi'=\varphi''=0$ yields optical theorem
\begin{eqnarray}
\sqrt{\frac{2\pi}{k}}\biggl\{2\cos(\mu\pi){\rm Im}\,f_c(k,\,0)-{\rm i}\sin(\mu\pi)
\int\limits_{-\pi}^{\pi}d\varphi[\Gamma^{(\nu)}(-\varphi)f_c(k,\,\varphi)+\biggr. \nonumber \\  \biggl.+
\Gamma^{(\nu)}(\varphi)f_c^*(k,\,\varphi)]\biggr\}
=\int\limits_{-\pi}^{\pi}d\varphi|f_c(k,\,\varphi)|^2.\label{eq40}
\end{eqnarray}
As the wavelength decreases, $f_0$ is decreasing as $O(k^{-1/2})$,
see (29), becoming negligible as compared to $f_c$. Thus, the
right-hand side of (40) tends to the total cross section in the
short-wavelength limit. Further, estimating appropriately the
integral in the left-hand side of (39), we find that this relation
in the short-wavelength limit turns out to be
\begin{eqnarray}
\frac 1{\rm i}\sqrt{\frac{k}{2\pi}}\cos(\mu\pi)\left[f_c(k,\,\varphi'-\varphi'')-
f^*_c(k,\,\varphi''-\varphi')\right]+2\sin^2(\mu\pi)\Delta^{(\nu)}_{kr_c}(\varphi'-\varphi'')+
\nonumber \\  +O(\sqrt{kr_c})=\frac{k}{2\pi}\int\limits_{-\pi}^{\pi}d\varphi\,f_c^*(k,\,\varphi-\varphi')f_c(k,\,\varphi-\varphi''),\label{eq41}
\end{eqnarray}
and the optical theorem in this limit takes form
\begin{equation}
2\sqrt{\frac{2\pi}{k}}\cos(\mu\pi){\rm Im}\,f_c(k,\,0)+\frac{4\pi}{k}\sin^2(\mu\pi)\Delta^{(\nu)}_{kr_c}(0)
+O(k^{-1})=\sigma,\label{eq42}
\end{equation}
where
\begin{equation}
\sigma=\int\limits_{-\pi}^{\pi}d\varphi |f_c(k,\,\varphi)|^2,\qquad kr_c\gg 1.\label{eq43}
\end{equation}

\subsection{Alternative decomposition of $S$-matrix}

The $S$-matrix can be presented as
\begin{equation}
S(k,\varphi,k_z;k',\varphi',k_z')=\left[I(k,\varphi;k',\varphi')+\delta(k-k')\frac{i}{\sqrt{2\pi k}}
\tilde{f}(k,\varphi-\varphi')\right]\delta(k_z-k_z'),\label{eq44}
\end{equation}
where $I(k,\,\varphi;\,k',\,\varphi')$ is the usual unity matrix
given by (15), and
\begin{equation}
\tilde{f}(k,\varphi)={\rm i}\sqrt{\frac{2\pi}{k}}[1-\cos(\mu\pi)]\Delta(\varphi)+f(k,\varphi);\label{eq45}
\end{equation}
then the asymptotics of the wave function can be rewritten as, cf.
(31),
\begin{equation}
\psi_{\bf k}({\bf r})=\psi_{\bf k}^{(0)}({\bf r})+\tilde{f}(k,\varphi)\frac{e^{{\rm i}(kr+\pi/4)}}{\sqrt{r}}
+O(r^{-3/2}).\label{eq46}
\end{equation}
Such a decomposition was proposed in \cite{Rui}, and the arguments
in its favour were given, for instance, in \cite{Sak}. Although the
unitarity condition in terms of amplitude $\tilde{f}$ takes the
conventional form given by (18), the question is whether it relates
physically meaningful (and thus finite) quantities. The answer is
clearly negative.

Indeed, one can easily get
\begin{eqnarray}
\frac{k}{2\pi}\int\limits_{-\pi}^{\pi}d\varphi\tilde{f}^*(k,\varphi-\varphi')\tilde{f}(k,\varphi-\varphi'')=
2[1-\cos(\mu\pi)]\Delta(\varphi'-\varphi'')-\nonumber \\-{\rm i}\sqrt{\frac{k}{2\pi}}
\left[f_c(k,\varphi'-\varphi'')-f^*_c(k,\varphi''-\varphi')\right],\label{eq47}
\end{eqnarray}
and, consequently, the total cross section in this framework,
\begin{equation}
\tilde{\sigma}=\int\limits_{-\pi}^{\pi}d\varphi|\tilde{f}(k,\varphi)|^2,\label{eq48}
\end{equation}
contains an unremovable divergency (actual infinity),
$\frac{4\pi}{k}[1-\cos(\mu\pi)]\Delta(0)$, for all wavelengths.

On the contrary, in the present paper we argue for a more physically
plausible framework. The total cross section in the long-wavelength
limit contains unremovable divergency,
$\frac{2\pi}{k}\sin^2(\mu\pi)\Delta(0)$ (see (38)), and this
divergency is due to an unphysical idealization involving singular
vortex. However, the total cross section in the short-wavelength
limit is given by (43) which is finite and, as it will be shown in
the following, independent of both flux ($\mu$) and boundary
($\rho$) parameters.

\subsection{Scattering amplitude}

The scattering amplitude in the short-wavelength limit has been
obtained in \cite{Si10} for the case of the Dirichlet boundary
condition. In Appendix A it is evaluated for the case of the Robin
boundary condition:
\begin{eqnarray}
f_c(k,\,\varphi)={\rm i}\sqrt{\frac{2\pi}{k}}\left[\cos(\mu\pi)
\Delta^{(\nu)}_{kr_c}(\varphi)-\sin(\mu\pi)\Gamma^{(\nu)}_{kr_c}(\varphi)\right]-
\sqrt{\frac{r_c}{2}|\sin(\varphi/2)|}\times  \nonumber \\
\nonumber \\ \times\exp\left\{-2{\rm i}kr_c|\sin(\varphi/2)| +{\rm
i}\mu[\varphi-{\rm sgn}(\varphi)\pi]\right\} \exp\left\{-{\rm
i}\left[2\chi^{(\rho)}(kr_c,\varphi)+\pi/4\right]\right\}+\nonumber \\
+\sqrt{r_c}O\left[(kr_c)^{-1/6}\right],\,\,
kr_c\gg 1,\label{eq49}
\end{eqnarray}
where
\begin{equation}
\chi^{(\rho)}(kr_c,\varphi)={\rm arctan}\left[\frac{2kr_c|\sin^3(\varphi/2)|}
{2{\rm cot}(\rho\pi)\sin^2(\varphi/2)-1}\right].\label{eq50}
\end{equation}
The Fraunhofer diffraction on the vortex in the forward direction is
described by the first term, while the classical reflection from the
vortex in other directions is described by the second term which,
apart from the phase factor, is the same as the second term in (20).
The explicit form of $\Delta^{(\nu)}_{kr_c}(\varphi)$ and
$\Gamma^{(\nu)}_{kr_c}(\varphi)$ is as follows:
\begin{equation}
\Delta^{(\nu)}_{kr_c}(\varphi)=\frac{e^{{\rm i}(\nu+\frac 12)\varphi}}{2\pi}\,\frac{\sin(s_c\varphi)}
{\sin(\varphi/2)}\label{eq51}
\end{equation}
and
\begin{equation}
\Gamma^{(\nu)}_{kr_c}(\varphi)=\frac{e^{{\rm i}(\nu+\frac 12)\varphi}}{2\pi}\,\frac{1-\cos(s_c\varphi)}
{\sin(\varphi/2)}\label{eq52}
\end{equation}
in the case
\begin{equation}
[\![kr_c+\mu]\!]-\nu=[\![kr_c-\mu]\!]+\nu+1=s_c,\label{eq53}
\end{equation}
or
\begin{equation}
\Delta^{(\nu)}_{kr_c}(\varphi)=\frac{e^{{\rm i}(\nu+\frac 12\mp \frac 12)\varphi}}{2\pi}\,
\frac{\sin\left[\left(s_c+\frac 12\right)\varphi\right]}{\sin(\varphi/2)}\label{eq54}
\end{equation}
and
\begin{equation}
\Gamma^{(\nu)}_{kr_c}(\varphi)=\frac{e^{{\rm i}(\nu+\frac 12\mp \frac 12)\varphi}}{2\pi}\,
\left\{\frac{1-\cos\left[\left(s_c+\frac 12\right)\varphi\right]}{\sin(\varphi/2)}
-{\rm tg}(\varphi/4)\pm{\rm i}\right\}\label{eq55}
\end{equation}
in the case
\begin{equation}
[\![kr_c+\mu]\!]-\nu-\frac 12\pm\frac 12=[\![kr_c-\mu]\!]+\nu+\frac 12\mp\frac 12=s_c.\label{eq56}
\end{equation}
Thus, in the short-wavelength limit, we get in the strictly forward
direction:
\begin{equation}
f_c(k,\,0)={\rm i}\sqrt{\frac{2k}{\pi}} \, r_c \cos(\mu\pi) + O(k^{-1/2}).\label{eq57}
\end{equation}

It should be noted that the left-hand side of (42) in the
nonvanishing order involves the contribution of the diffraction peak
only, whereas the right-hand side of (42) includes the contribution
of the classical reflection as well. In Appendix B we show that the
total cross section in the short-wavelength limit is independent of
the vortex flux, as well as of the choice of the boundary condition
from the variety of the Robin ones:
\begin{equation}
\sigma=4r_c+O(k^{-1}),\label{eq58}
\end{equation}
this is twice the classical total cross section, the latter been
equal to $2r_c$. Already the differential cross section in the
short-wavelength limit is independent of the choice of the boundary
condition:
\begin{eqnarray}
\frac{d\sigma}{d\varphi}=|f_c(k,\varphi)|^2=2r_c\left\{\cos(2\mu\pi)\tilde{\Delta}_{kr_c}(\varphi)
+\right.\nonumber\\ \left.+\!\left[1\!-\!\cos(2\mu\pi)\!-\!\sin(2\mu\pi)\sin(kr_c\varphi)\right]
\tilde{\Delta}_{\frac12kr_c}(\varphi)\!\right\}
\!+\!\frac{r_c}{2}|\sin(\varphi/2)|\!+\!r_cO\left[
(kr_c)^{\!-1/2}\right],\label{eq59}
\end{eqnarray}
where
\begin{equation}
\tilde{\Delta}_x(\varphi)=\frac{1}{4\pi x}\frac{\sin^2(x\varphi)}{\sin^2(\varphi/2)}\qquad
(x\gg 1,\,-\pi<\varphi<\pi)\label{eq60}
\end{equation}
is a one more regularization of the angular delta-function,
satisfying (12). By integrating (59) over all angles, one gets
easily (58). A third way to get (58) is to insert (57) into optical
theorem (42). Thus, the contribution of the diffraction peak to the
total cross section is flux independent and is equal to that of
classical reflection. The latter is well known for scattering off a
tube with zero magnetic flux, see, e.g., \cite{Mor}, and, for the
case of nonzero magnetic flux, it has been first established under
the Dirichlet boundary condition \cite{Si10}.

It is instructive to derive the explicit form of
$\Gamma^{(\nu)}(\varphi)$ (24) here. Using elementary trigonometric
relation
$$
{\rm cot}(\varphi/2)\left\{\sin[(n+1)\varphi]-\sin(n\varphi)\right\}=\cos[(n+1)\varphi]+\cos(n\varphi),
$$
one can get
$$
\int\limits_{0}^{\pi}d\varphi\,{\rm cot}(\varphi/2)\sin(N\varphi)=\pi,\qquad N=1,\,2,\,3,\,\ldots,
$$
which results in relation
$$
{\rm cot}(\varphi/2)=2\sum\limits_{n\in\mathbb{Z}\atop{n\geq1}}\sin(n\varphi).
$$
The use of the latter along with the definition (5) yields
\begin{equation}
\sum\limits_{n\in\mathbb{Z}\atop n\geq N}e^{{\rm i}n\varphi}=
\pi\Delta(\varphi)-e^{{\rm i}N\varphi}(e^{{\rm i}\varphi}-1)^{-1},\quad N=1,\,2,\,3,\,\ldots,\label{eq61}
\end{equation}
whence it follows that
\begin{equation}
\Gamma^{(\nu)}(\varphi)=\frac{e^{{\rm i}(\nu+\frac 12)\varphi}}{2\pi}\frac{1}{\sin(\varphi/2)},\label{eq62}
\end{equation}
where the divergence at $\varphi=2\pi l$ ($l\in \mathbb{Z}$) in
(62), as well as in the second term in (61), is to be understood in
the principal-value sense.

Although amplitude $f_0$ (29) with $\Gamma^{(\nu)}$ (62) diverges in
the forward direction, this divergence has no physical consequences,
because there is a crossover to another regime in the strictly
forward direction: amplitude $f_0$, instead of being proportional to
$k^{-1/2}$, becomes, formally, proportional to $r^{1/2}$. This is
most easily seen from the following expression which is valid for
all scattering angles and has been obtained in \cite{Som}:
\begin{eqnarray}
f_0(k,\varphi)\frac{e^{{\rm i}(kr+\pi/4)}}{\sqrt{r}}={\rm i}\sin(\mu\pi)
e^{{\rm i}kr\cos\varphi}e^{{\rm i}(\nu+\frac 12)\varphi}\times \nonumber \\
\times{\rm sgn}[\sin(\varphi/2)]{\rm erfc}\left[e^{-{\rm i}\pi/4}\sqrt{2kr}|\sin(\varphi/2)|\right],\label{eq63}
\end{eqnarray}
where ${\rm erfc}(z)=\frac{2}{\sqrt
\pi}\int\limits_{z}^{\infty}du\,e^{-u^2}$ is the complementary error
function. In the strictly forward direction one gets a
discontinuity,
\begin{equation}
\lim\limits_{\varphi\rightarrow\pm0}f_0(k,\varphi)\frac{e^{{\rm i}(kr+\pi/4)}}{\sqrt{r}}=
\pm{\rm i}\sin(\mu\pi)e^{{\rm i}kr},\label{eq64}
\end{equation}
which cancels the discontinuity of the incident wave (first term in
(31)), see \cite{Ola,Afa,Si5,Ste,Alv}. Consequently, wave function
(31) is finite and continuous in the forward direction:
\begin{equation}
\psi_{\bf k}({\bf r})|_{\varphi=0}=\cos(\mu\pi)e^{{\rm i}kr}+f_c(k,\,0)\frac{e^{{\rm i}(kr+\pi/4)}}{\sqrt{r}},\label{eq65}
\end{equation}
that is consistent with its exact expression (33). The appearance of
factor $\cos(\mu\pi)$ in the transmitted wave (first term in (65))
can be intuitively understood as a result of self-interference from
different sides of the vortex \cite{Ste, Alv}. As it is shown in the
present paper, the same factor appears also in the scattered wave
(second term in (65)) due to the Fraunhofer diffraction, see (57).

\section{Summary and discussion}

The $S$-matrix, its unitarity and the consequent optical theorem for
the case of the Aharonov-Bohm scattering are considered in the
present paper. Standard scattering theory (see, e.g., \cite{Ree}) is
not applicable here, since the interaction with scatterer is not of
short-range type. That is why, although the $S$-matrix is evidently
unitary, its unitarity condition in terms of the scattering
amplitude takes a rather unusual form.

In the ultraquantum (long-wavelength, $k\rightarrow0$) limit when
the vortex thickness is neglected, the unitarity condition takes
form (38) with no terms which are linear in the scattering
amplitude. The scattering amplitude in the ultraquantum limit, see
(29) with (62), was first obtained by Aharonov and Bohm \cite{Aha}
and then rederived in the framework of different approaches
(perhaps, the one with the use of $\Gamma^{(\nu)}(\varphi)$ (24),
presented here, is the most simple and straightforward). As to the
behaviour of this amplitude in the forward direction, the only thing
that should be borne in mind is that the divergence has to be
understood in the principal-value sense. The total cross section in
the ultraquantum limit diverges as well, hence the optical theorem
in this limit seems to be hardly efficient, being merely a
consistency relation between two divergent (infinite) quantities,
$\Delta(0)$.

The divergence of the scattering amplitude and the total cross
section, as well as the failure with the optical theorem, has no
physical meaning, being an artefact of the approximation which
neglects the vortex thickness: this is certainly an excessive
idealization, whereas any realistic vortex is of finite nonzero
thickness. As long as one departs from the ultraquantum limit and
the vortex thickness ($2r_c$) is taken into account, the unitarity
condition can be formulated as relation (39) involving the
$r_c$-dependent part of the scattering amplitude, $f_c$ (34); the
optical theorem relates $f_c$ in the forward direction to the
contribution of $f_c$ to the total cross section, see (40).

In the quasiclassical (short-wavelength, $k\rightarrow \infty$)
limit, the vortex-thickness effects are prevailing and $f_c$
approximates fairly the whole scattering amplitude. The unitarity
condition in this limit takes form (41), and the optical theorem is
given by (42). The scattering amplitude in the quasiclassical limit,
$f_c$ (49), consists of two parts: the one corresponding to the
Fraunhofer diffraction on the vortex in the forward direction and
the other one corresponding to the classical reflection from the
vortex in all other directions. The latter, apart from the phase
factor, is flux independent, whereas the former is essentially flux
dependent being periodic in the value of the flux with the period
equal to the London flux quantum. We conclude that the persistence
of the Fraunhofer diffraction in the short-wavelength limit is
crucial for maintaining the optical theorem in the quasiclassical
limit, since the classical reflection vanishes in the forward
direction.

It should be emphasized that, although only the contribution of the
diffraction peak is involved in the left-hand side of (42) in the
nonvanishing order, the right-hand side of (42) includes both the
contributions of the diffraction peak and the classical reflection.
Separate flux dependent terms in the left-hand side of (42)
compensate each other to yield the flux independent right-hand side
of (42), which equals the doubled diameter of the vortex, i.e. the
doubled classical total cross section.

Thus, quantum-mechanical scattering theory in the quasiclassical
limit gives an effect which is alien to classical mechanics: a
particle wave penetrates in the forward direction behind an
obstacle, and the Fraunhofer diffraction persisting in the
short-wavelength limit is an essential ingredient of this effect.
The $r_c$-independent part of the particle wave function (see either
(7) or (33)) produces a transmitted wave, while the $r_c$-dependent
part produces a scattered wave which is due to the Fraunhofer
diffraction. If the obstacle is an impermeable magnetic vortex, then
both the transmitted and the scattered waves are modulated by cosine
of the vortex flux in the units of the London flux quantum, see (65)
and (57). All this comprises the scattering Aharonov-Bohm effect
persisting in the quasiclassical limit \cite{Si10}. As it is shown
in the present paper, this phenomenon is the same for the choice of
either Dirichlet boundary condition (6), or Neumann one (36), and,
in general, is independent of the choice of boundary conditions from
the variety of Robin ones (32). It would be interesting to observe
this phenomenon directly in a scattering experiment with
short-wavelength, almost classical, particles.

\section*{Acknowledgments}

The work was partially supported by the Department of Physics and
Astronomy of the National Academy of Sciences of Ukraine under
special program ``Fundamental properties of physical systems in
extremal conditions''.

\renewcommand{\thesection}{A}
\renewcommand{\theequation}{\thesection.\arabic{equation}}
\setcounter{section}{1} \setcounter{equation}{0}

\section*{Appendix A}

Let us denote the infinite sum (with factor 2) which enters
amplitude $f_c$ (34) by
\begin{equation}
\Sigma(s,\varphi)=2\sum\limits_{n\in\mathbb{Z}}e^{{\rm i}n\varphi}e^{{\rm i}(|n|-|n-\mu|)\pi}
\Upsilon_{|n-\mu|}^{(\rho)}(s),
\end{equation}
where $s=kr_c$, and decompose it into three parts
\begin{equation}
\Sigma(s,\varphi)=\Sigma_1(s,\varphi)+\Sigma_2(s,\varphi)+\Sigma_3(s,\varphi),
\end{equation}
where
\begin{equation}
\Sigma_1(s,\varphi)=\sum\limits_{|n-\mu|\leq s}e^{{\rm i}n\varphi}e^{{\rm i}(|n|-|n-\mu|)\pi},
\end{equation}
\begin{equation}
\Sigma_2(s,\varphi)=\sum\limits_{|n-\mu|\leq s}e^{{\rm i}n\varphi}e^{{\rm i}(|n|-|n-\mu|)\pi}
\frac{H_{|n-\mu|}^{(2)}(s)}{H_{|n-\mu|}^{(1)}(s)}\,\frac{\cot(\rho\pi)+s\frac{d}{ds}\ln H_{|n-\mu|}^{(2)}(s)}
{\cot(\rho\pi)+s\frac{d}{ds}\ln H_{|n-\mu|}^{(1)}(s)},
\end{equation}
\begin{equation}
\Sigma_3(s,\varphi)=2\sum\limits_{|n-\mu|>s}e^{{\rm i}n\varphi}e^{{\rm i}(|n|-|n-\mu|)\pi}
\frac{J_{|n-\mu|}(s)}{H_{|n-\mu|}^{(1)}(s)}\,\frac{\cot(\rho\pi)+s\frac{d}{ds}\ln J_{|n-\mu|}(s)}
{\cot(\rho\pi)+s\frac{d}{ds}\ln H_{|n-\mu|}^{(1)}(s)}.
\end{equation}
The first part is easily calculated as a finite sum of geometric
progression, yielding
\begin{equation}
\Sigma_1(s,\varphi)=2\pi\left[\cos(\mu\pi)\Delta_s^{(\nu)}(\varphi)-\sin(\mu\pi)\Gamma_s^{(\nu)}(\varphi)\right],
\end{equation}
with $\Delta_s^{(\nu)}(\varphi)$ and $\Gamma_s^{(\nu)}(\varphi)$
given by (51)-(56).

The second part is presented as
\begin{eqnarray}
\Sigma_2(s,\varphi)=\sum\limits_{n=\nu+1}^{s_+}e^{{\rm i}(n\varphi+\mu\pi)}
\frac{H_{n-\mu}^{(2)}(s)}{H_{n-\mu}^{(1)}(s)}\,\frac{\cot(\rho\pi)+s\frac{d}{ds}\ln H_{n-\mu}^{(2)}(s)}
{\cot(\rho\pi)+s\frac{d}{ds}\ln H_{n-\mu}^{(1)}(s)}+\nonumber \\
+\sum\limits_{n=-\nu}^{s_-}e^{-{\rm i}(n\varphi+\mu\pi)}
\frac{H_{n+\mu}^{(2)}(s)}{H_{n+\mu}^{(1)}(s)}\,\frac{\cot(\rho\pi)+s\frac{d}{ds}\ln H_{n+\mu}^{(2)}(s)}
{\cot(\rho\pi)+s\frac{d}{ds}\ln H_{n+\mu}^{(1)}(s)},
\end{eqnarray}
where $s_\pm=[\![s\pm\mu]\!]$. Using the asymptotics of the Hankel
functions at $\alpha<s_{\rm min}\leq s<\infty$ (see, e.g.,
\cite{Abra}):
\begin{eqnarray}
H_\alpha^{(1)}(s)\!=\!\left(\frac{2e^{-{\rm i}\pi/2}}{\pi}\right)^{1/2}\!\left(s^2\!-\!\alpha^2\right)^{-1/4}
\exp\!\left[{\rm i}\left(s^2\!-\!\alpha^2\right)^{1/2}\!-\!{\rm i}\alpha\arccos\left(\alpha/s\right)\right] \times
\nonumber \\ \times
\!\left[1\!+\!O(s^{-1})\right],
\end{eqnarray}
\begin{eqnarray}
H_\alpha^{(2)}(s)\!=\!\left(\frac{2e^{{\rm i}\pi/2}}{\pi}\right)^{1/2}\!\left(s^2\!-\!\alpha^2\right)^{-1/4}
\exp\!\left[-{\rm i}\left(s^2\!-\!\alpha^2\right)^{1/2}\!+\!{\rm i}\alpha\arccos\left(\alpha/s\right)\right]\times
\nonumber \\  \times
\!\left[1\!+\!O(s^{-1})\right],
\end{eqnarray}
we get
\begin{equation}
\Sigma_2(s,\varphi)={\rm i}\sum\limits_{n=\nu+1}^{s_+}e^{2\pi{\rm i}f_+(n,s)}+
{\rm i}\sum\limits_{n=-\nu}^{s_-}e^{2\pi{\rm i}f_-(n,s)}+O(s^{-1}),
\end{equation}
where
\begin{eqnarray}
f_\pm(n,s)=\pi^{-1}\Biggl\{-\left[s^2-(n\mp \mu)^2\right]^{1/2}+(n\mp\mu)\arccos\left(
\frac{n\mp\mu}{s}\right)-\Biggr. \nonumber \\
\Biggl.-\arctan\left(\frac{\left[s^2-(n\mp\mu)^2\right]^{1/2}}{\cot(\rho\pi)-\frac{s^2}{2}
\left[s^2-(n\mp\mu)^2\right]^{-1}}\right)\pm\frac 12n\varphi\pm\frac 12\mu\pi\Biggr\}.
\end{eqnarray}

Using an expansion into the Fourier series, one can get relation
\begin{eqnarray}
\sum\limits_{n=n_1}^{n_2}\exp[2\pi{\rm i}f(n)]=\sum\limits_{l\in\mathbb{Z}}\int\limits_{n_1}^{n_2}du\,
\exp\{2\pi{\rm i}[f(u)-lu]\}+ \nonumber \\
+\frac 12\exp[2\pi{\rm i}f(n_1)]+\frac 12\exp[2\pi{\rm i}f(n_2)].
\end{eqnarray}
If $n_2-n_1\gg1$ and $f(u)$ is convex upwards on interval
$n_1<u<n_2$, $\frac{d^2f}{du^2}<0$, then only a finite number of
terms in the series on the right-hand side of (A.12) makes the main
contribution, and one can use the method of stationary phase for its
evaluation. Namely, let $s\gg1$, $h(u)$ be a continuous function,
and $g(u)$ be a real function that is convex upwards,
$\frac{d^2g}{du^2}<0$, with one stationary point,
$\left.\frac{dg}{du}\right|_{u=u_0}=0$, inside the interval,
$n_1<u_0<n_2$. Then
\begin{equation}
\int\limits_{n_1}^{n_2}du\,h(u)\exp[{\rm i}sg(u)]=h(u_0)\exp[{\rm i}sg(u_0)]
\left[\frac{2\pi e^{{\rm i}\pi/2}}{s(d^2g/du^2)|_{u=u_0}}\right]^{1/2}+O(s^{-1}).
\end{equation}

In our case
\begin{equation}
h_\pm(u)=\exp\left\{-2{\rm i}\arctan\left(\frac{\left[s^2-(u\mp\mu)^2\right]^{1/2}}
{\cot(\rho\pi)-\frac{s^2}{2}\left[s^2-(u\mp\mu)^2\right]^{-1}}\right)\right\},
\end{equation}
\begin{equation}
g_\pm^{(l)}(u)\!=\!\frac 2s\left\{-\left[s^2\!-\!(u\mp\mu)^2\right]^{1/2}\!+\!(u\mp\mu)
\arccos\left(\frac{u\mp\mu}{s}\right)\pm\frac 12u\varphi\pm\frac 12\mu\pi\!-\!ul\pi\right\},
\end{equation}
and one gets
\begin{equation}
\frac{d}{du}g_{\pm}^{(l)}=\frac 2s\left[\arccos\left(\frac{u\mp\mu}{s}\right)\pm\frac 12\varphi-l\pi\right],
\end{equation}
\begin{equation}
\frac{d^2}{du^2}g_{\pm}^{(l)}=-\frac 2s\left[s^2-(u\mp\mu)^2\right]^{-1/2}.
\end{equation}
Thus, $g_+^{(l)}(u)$ is convex upwards on interval $\nu+1<u<s_+$,
and $g_-^{(l)}(u)$ is convex upwards on interval $-\nu<u<s_-$.
Taking $\varphi$ from the range $-\pi<\varphi<\pi$, one finds that,
since $0<u\mp\mu<s$, the stationary point inside the interval exists
only at $l=0$ and is determined by
\begin{equation}
\arccos\left(\frac{u_0-\mu}{s}\right)=-\frac 12\varphi \qquad {\rm at} \qquad -\pi<\varphi<0
\end{equation}
for $g_+^{(0)}(u)$ and by
\begin{equation}
\arccos\left(\frac{u_0+\mu}{s}\right)=\frac 12\varphi \qquad {\rm at} \qquad 0<\varphi<\pi
\end{equation}
for $g_-^{(0)}(u)$; the integrals with $l\neq 0$ make a contribution
of order $O(1)$. Consequently, we get
\begin{eqnarray}
\sum\limits_{n=\nu+1}^{s_+}e^{2\pi{\rm i}f_+(n,s)}=\left[-s\pi e^{{\rm i}\pi/2}\sin(-\varphi/2)\right]^{1/2}\times
\nonumber \\
\times \exp\left\{\!-2{\rm i}\arctan\left[\frac{2s\sin^3(-\varphi/2)}
{2\cot(\rho\pi)\sin^2(\!-\varphi/2)\!-\!1}\right]\!\right\}
\exp\left[\!-2{\rm i}s\sin(\!-\varphi/2)\!+\!{\rm i}\mu(\varphi+\pi)\right]\!+
\nonumber \\+O(1),\quad -\pi<\varphi<0 \nonumber \\
\end{eqnarray}
and
\begin{eqnarray}
\sum\limits_{n=-\nu}^{s_-}e^{2\pi{\rm i}f_-(n,s)}=\left[-s\pi e^{{\rm i}\pi/2}\sin(\varphi/2)\right]^{1/2}\times
\nonumber \\
\times \exp\left\{\!-2{\rm i}\arctan\left[\frac{2s\sin^3(\varphi/2)}
{2\cot(\rho\pi)\sin^2(\varphi/2)\!-\!1}\right]\!\right\}
\exp\left[\!-2{\rm i}s\sin(\varphi/2)\!+\!{\rm i}\mu(\varphi-\pi)\right]\!+
\nonumber \\+O(1),\quad 0<\varphi<\pi. \nonumber \\
\end{eqnarray}
Combining (A.20) and (A.21), we obtain
\begin{eqnarray}
\Sigma_2(s,\varphi)=\left[s\pi e^{{\rm i}\pi/2}|\sin(\varphi/2)|\right]^{1/2}\exp\left[-2{\rm i}\chi^{(\rho)}
(s,\varphi)\right]\times
\nonumber \\
\times \exp\left\{-2{\rm i}s|\sin(\varphi/2)|+{\rm i}\mu\left[\varphi-{\rm sgn}(\varphi)\pi\right]\right\}+
O(1),\quad -\pi<\varphi<\pi,
\end{eqnarray}
where $\chi^{(\rho)}(s,\varphi)$ is given by (50).

To evaluate the last part in (A.2), we rewrite it as
\begin{eqnarray}
\Sigma_3(s,\varphi)=2\sum\limits_{n=s_++1}^{\infty}e^{{\rm i}(n\varphi+\mu\pi)}
\frac{J_{n-\mu}(s)}{H_{n-\mu}^{(1)}(s)}\,\frac{\cot(\rho\pi)+s\frac{d}{ds}\ln J_{n-\mu}(s)}
{\cot(\rho\pi)+s\frac{d}{ds}\ln H_{n-\mu}^{(1)}(s)}+ \nonumber \\
+2\sum\limits_{n=s_-+1}^{\infty}e^{-{\rm i}(n\varphi+\mu\pi)}
\frac{J_{n+\mu}(s)}{H_{n+\mu}^{(1)}(s)}\,\frac{\cot(\rho\pi)+s\frac{d}{ds}\ln J_{n+\mu}(s)}
{\cot(\rho\pi)+s\frac{d}{ds}\ln H_{n+\mu}^{(1)}(s)},
\end{eqnarray}
and use the asymptotics of the cylindrical functions at $s\leq
s_{\rm max}<\alpha$ \cite{Abra}:
\begin{equation}
J_\alpha(s)=(2\pi)^{-1/2}(\alpha^2-s^2)^{-1/4}\exp\left[\sqrt{\alpha^2-s^2}-\alpha\,{\rm arccosh}
\left(\alpha/s\right)\right]\left[1+O(\alpha^{-1})\right],
\end{equation}
\begin{equation}
Y_\alpha(s)=-\left(\frac 2\pi\right)^{1/2}(\alpha^2-s^2)^{-1/4}\exp\left[-\sqrt{\alpha^2-s^2}+\alpha\,{\rm arccosh}
\left(\alpha/s\right)\right]\left[1+O(\alpha^{-1})\right],
\end{equation}
where $Y_\alpha(s)$ is the Neumann function of order $\alpha$
($H_\alpha^{(1)}(s)=J_\alpha(s)+{\rm i}Y_\alpha(s)$). We get
\begin{equation}
\Sigma_3(s,\varphi)={\rm i}\Sigma_3^{(+)}(s,\varphi)+{\rm i\Sigma_3^{(-)}}(s,\varphi),
\end{equation}
where
\begin{eqnarray}
\Sigma_3^{(\pm)}(s,\varphi)\!=\!\sum\limits_{n=s_\pm+1}^{\infty}\exp\!\left[
2\sqrt{(n\!\mp\!\mu)^2\!-\!s^2}-\!2(n\mp\mu)\,{\rm arccosh}\left(\frac{n\mp\mu}{s}\right)\right]\!\times
\nonumber \\ \times
\left[1+F^{(\rho)}\left(\frac{n\mp\mu}{s},s\right)\right]^{-1}e^{\pm{\rm i}(n\varphi+\mu\pi)}
\left\{1+O\left[(n\mp\mu)^{-1}\right]\right\},
\end{eqnarray}
\begin{equation}
F^{(\rho)}(v,s)=\left[e^{2{\rm i}\kappa(v,s)}-1\right]s\sqrt{v^2-1}\left[
\cot(\rho\pi)+\frac 12(v^2-1)^{-1}+s\sqrt{v^2-1}\right]^{-1},
\end{equation}
\begin{equation}
\kappa(v,s)=\arctan\left\{2\exp\left[2s\left(v\,{\rm
arccosh}v-\sqrt{v^2-1}\right)\right]\right\}.
\end{equation}
We evaluate sum (A.27) at $\varphi=0$ as
\begin{eqnarray}
\Sigma_3^{(\pm)}(s,0)&\!\!=e^{\pm{\rm i}\mu\pi}\int\limits_{s_\pm+1}^{\infty}du\,
\exp\left[2\sqrt{(u\mp\mu)^2-s^2}-2(u\mp\mu)\,{\rm arccosh}\left(\frac{u\mp\mu}{s}\right)\right]\times \nonumber \\
&\times\left[1+F^{(\rho)}\left(\frac{u\mp\mu}{s},s\right)\right]^{-1}+O(1),
\end{eqnarray}
since the real part of the logarithmic derivative of the integrand
in (A.30) is negative for $s_\pm+1<u<\infty$. Further, by
substituting $u\mp\mu=s\cosh \tau$, we get
\begin{eqnarray}
e^{\mp{\rm i}\mu\pi}\Sigma_3^{(\pm)}(s,0)=s\int\limits_{0}^{\infty}d\tau\,
\sinh\tau\exp\left[2s(\sinh\tau-\tau\cosh\tau)\right]\times \nonumber \\
\times\left[1+F^{(\rho)}(\cosh\tau,s)\right]^{-1}+O(1).
\end{eqnarray}
The integral in (A.31) is estimated with the use of the Laplace
method, yielding
\begin{equation}
e^{\mp{\rm i}\mu\pi}\Sigma_3^{(\pm)}(s,0)=\Gamma\left(\frac 23\right)\left(
\frac{s}{12}\right)^{1/3}+O(1)
\end{equation}
at $s\rightarrow\infty$, where $\Gamma(u)$ is the Euler
gamma-function. Using the last relation, we get estimate
\begin{equation}
\left|\Sigma_3(s,\varphi)\right|\leq cs^{1/3},
\end{equation}
where constant $c$ is independent of $s$ and $\varphi$.

Collecting (A.6), (A.22) and (A.33), and multiplying by ${\rm
i}(2\pi k)^{-1/2}$, we obtain (49).

\renewcommand{\thesection}{B}
\renewcommand{\theequation}{\thesection.\arabic{equation}}
\setcounter{section}{1} \setcounter{equation}{0}
\section*{Appendix B}

Let us consider the right-hand side of (40):
\begin{eqnarray}
\int\limits_{-\pi}^{\pi}d\varphi|f_c(k,\varphi)|^2=\frac 4k\sum\limits_{n\in \mathbb{Z}}
\left[\Upsilon_{|n-\mu|}^{(\rho)}(kr_c)\right]^*\Upsilon_{|n-\mu|}^{(\rho)}(kr_c)=\nonumber \\
\left.=\!\!\frac 4k\!\sum\limits_{n\in\mathbb{Z}}\!\frac{\left\{\left[\cot(\rho\pi)+u\frac{d}{du}\right]
J_{|n-\mu|}(u)\right\}^2}{\left\{\!\left[\!\cot(\rho\pi)\!+\!u\frac{d}{du}\!\right]
\!J_{|n-\mu|}(u)\!\right\}^2\!+\!\left\{\!\left[\!\cot(\rho\pi)\!+\!u\frac{d}{du}\!\right]
\!Y_{|n-\mu|}(u)\right\}^2}\right|_{u=kr_c}. \nonumber \\
\end{eqnarray}
In the short-wavelength limit, the sum is cut at $|n-\mu|=kr_c$ (see
Appendix A), and the large-argument asymptotics for the cylindrical
functions is used. As a result, we get the following expression for
$\sigma$ (43):
\begin{eqnarray}
\sigma=\frac 4k
\left\{\left[\cot(\rho\pi)-\frac 12\right]^2+k^2r_c^2\right\}^{-1}\times \nonumber \\
\times\sum\limits_{|n-\mu|\leq kr_c}\left\{\left[\cot(\rho\pi)-\frac 12\right]\cos
\left(kr_c-\frac 12|n-\mu|\pi-\frac{\pi}{4}\right)-\right. \nonumber \\ \left.-kr_c
\sin\left(kr_c-\frac 12|n-\mu|\pi-\frac{\pi}{4}\right)\right\}^2.
\end{eqnarray}
Performing the summation, we get
\begin{equation}
\sigma=\frac{4\pi}{k}\left[\Delta_{kr_c}^{(\nu)}(0)+\cos(\mu\pi)\Delta_{kr_c}^{(\nu)}(\pi)
A_{1}^{(\rho)}(kr_c)+{\rm i}\sin(\mu\pi)\Gamma_{kr_c}^{(\nu)}(\pi)A_2^{(\rho)}(kr_c)\right],
\end{equation}
where
\begin{equation}
A_1^{(\rho)}(u)=\sin\left\{2u+2\arctan\left[\frac{2u}{2\cot(\rho\pi)-1}\right]\right\}
\end{equation}
and
\begin{equation}
A_2^{(\rho)}(u)=\cos\left\{2u+2\arctan\left[\frac{2u}{2\cot(\rho\pi)-1}\right]\right\}.
\end{equation}
Using (51)-(56), we get
\begin{equation}
\sigma=\frac{4}{k}s_c-\frac 2k\sin(\mu\pi)e^{{\rm i}\nu\pi}\left(1-e^{{\rm i}s_c\pi}\right)A_2^{(\rho)}(kr_c)
\end{equation}
in the case when $s_c$ is given by (53), or
\begin{equation}
\sigma=\frac{4}{k}(s_c+\frac 12)\pm\frac 2k\cos(\mu\pi)e^{{\rm i}\nu\pi}
e^{{\rm i}s_c\pi}A_1^{(\rho)}(kr_c)-\frac2k\sin(\mu\pi)e^{{\rm i}\nu\pi}A_2^{(\rho)}(kr_c)
\end{equation}
in the case when $s_c$ is given by (56). Since $s_c$ is estimated as
$s_c=kr_c+O(1)$ in the short-wavelength limit, we get (58).

\end{document}